RESEARCH ARTICLE

# Predicting 3D structure and stability of RNA pseudoknots in monovalent and divalent ion solutions

Ya-Zhou Shi[1,2☯], Lei Jin[2☯], Chen-Jie Feng[2], Ya-Lan Tan[2], Zhi-Jie Tan[2]*

**1** Research Center of Nonlinear Science, School of Mathematics and Computer Science, Wuhan Textile University, Wuhan, China, **2** Department of Physics and Key Laboratory of Artificial Micro- and Nano-structures of Ministry of Education, School of Physics and Technology, Wuhan University, Wuhan, China

☯ These authors contributed equally to this work.
* zjtan@whu.edu.cn

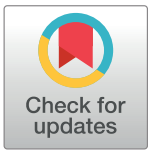

**Data Availability Statement:** All relevant computational data are within the paper and its Supporting Information files. Experimental data are publicly available from published papers cited in the work. Two web services MC-Fold/MC-Sym pipeline and RNAComposer used in the work are available by http://www.major.iric.ca/MC-Fold/ and http://rnacomposer.ibch.poznan.pl/, respectively.

**Funding:** This work was supported by the National Science Foundation of China grants [11774272 to ZJT, 11605125 to YZS, 11575128 and 11374234

## Abstract

RNA pseudoknots are a kind of minimal RNA tertiary structural motifs, and their three-dimensional (3D) structures and stability play essential roles in a variety of biological functions. Therefore, to predict 3D structures and stability of RNA pseudoknots is essential for understanding their functions. In the work, we employed our previously developed coarse-grained model with implicit salt to make extensive predictions and comprehensive analyses on the 3D structures and stability for RNA pseudoknots in monovalent/divalent ion solutions. The comparisons with available experimental data show that our model can successfully predict the 3D structures of RNA pseudoknots from their sequences, and can also make reliable predictions for the stability of RNA pseudoknots with different lengths and sequences over a wide range of monovalent/divalent ion concentrations. Furthermore, we made comprehensive analyses on the unfolding pathway for various RNA pseudoknots in ion solutions. Our analyses for extensive pseudoknots and the wide range of monovalent/divalent ion concentrations verify that the unfolding pathway of RNA pseudoknots is mainly dependent on the relative stability of unfolded intermediate states, and show that the unfolding pathway of RNA pseudoknots can be significantly modulated by their sequences and solution ion conditions.

## Author summary

RNA pseudoknotted structures and their stability can play important roles in RNA cellular functions such as transcription, splicing and translation. Due to the polyanionic nature of RNAs, metal ions such as $Na^+$ and $Mg^{2+}$ in solutions can play an essential role in RNA folding. Although several computational models have been developed to predict 3D structures for RNA pseudoknots to further unveil the mechanisms of their functions, these structure prediction models seldom consider ion conditions departing from the high salt (e.g., 1M NaCl) and temperatures from the room temperature. In this work, we employed our coarse-grained model to predict 3D structures and thermodynamic stability for to ZJT], and the Program for New Century Excellent Talents [Grant No. NCET 08-0408 to ZJT]. The funders had no role in study design, data collection and analysis, decision to publish, or preparation of the manuscript.

**Competing interests:** The authors have declared that no competing interests exist.

various RNA pseudoknots in monovalent/divalent ion solutions from their sequences, and made comparisons with extensive experimental data and existing models. In addition, based on our comprehensive analyses for extensive pseudoknots and the wide range of monovalent/divalent ion conditions, we confirmed that the thermally unfolding pathway of RNA pseudoknots is mainly determined by the relative stability of intermediate states, which has been proposed by Thirumalai et al. Our analyses also show that the thermally unfolding pathway of RNA pseudoknots could be apparently modulated by the sequences and ion conditions.

## Introduction

RNAs can fold into complex three-dimensional (3D) structures to carry out their various biological functions [1]. An RNA pseudoknot represents a very common structure motif, which is not only one of the fundamental structure elements in various classes of RNAs such as human telomerase RNA, self-splicing introns of ribozyme and S-adenosylmethionine-responsive riboswitches, but also involved in many biological functions, including regulation and catalysis [2,3]. For instance, an RNA pseudoknot can be present within the coding regions of an mRNA, where it stimulates programmed -1 ribosomal frameshifting to control the relative expression levels of proteins [2±4]. Generally, an RNA pseudoknot is formed when a sequence of nucleotides within a single-stranded loop region forms base pairs with a complementary sequence outside that loop [2,3,5,6]. Many experiments have shown that this special 3D topology is key to realize the various functions of RNA pseudoknots [2±4]. In addition, the stability of RNA pseudoknots can also play important roles in modulating their biological functions, and structure changes of RNA pseudoknots could cause diseases such as dyskeratosis [3,7,8]. Thus, to determine 3D structures and quantify stability of RNA pseudoknots is essential to unveil the mechanisms of their functions and to further aid the related drug design [5,9].

There have been several successful experimental methods to obtain 3D structures of RNAs, such as X-ray crystallography, nuclear magnetic resonance spectroscopy, and newly developed cryo-electron microscopy [9±12]. However, it is still very time-consuming and expensive to derive high-resolution 3D structures of RNAs and the RNA structures deposited in Protein Data Bank (PDB) are still limited [9,12]. To complement experimental measurements, some computational models have been developed to predict 3D structures for RNAs [13±22]. The knowledge-based models [23±34] such as MC-Fold/MC-Sym pipeline [24], FARNA [25], 3dRNA [29,35,36], RNAComposer [30] and pk3D [31] are rather successful and efficient in constructing 3D structures for RNA pseudoknots through fragments assembly based on limited experimental structures/fragments or reliable secondary structures, while it is still a problem to exactly predict secondary structures of RNA pseudoknots [11,20]. Furthermore, most of the above methods cannot give reliable predictions for the thermodynamic properties of RNA pseudoknots from their sequences [9±11].

Simultaneously, some coarse-grained (CG) models have been developed to predict the thermodynamic stability of RNAs including pseudoknots [37±46]. The Vfold model enables predictions for the structure, stability, and the free energy landscape for RNA pseudoknots from sequences through enumerating loop conformations on a diamond lattice [37,38]. The model is applicable to secondary structure folding while the 3D structures need to be built through fragment assembly based on secondary structures [47]. Several other CG models such as the iFoldRNA [39], the HiRE-RNA [40] and the oxRNA [42] have been used to predict 3D structure and stability for a few RNA pseudoknots, but the parameters of these models may need further validation for quantifying RNA thermodynamics to accord with experiments. In addition, due to the polyanionic nature of RNAs, metal ions (e.g., $Na^+$ and $Mg^{2+}$) in ion solutions can play an essential role in RNA folding [48±53], and $Mg^{2+}$ can play a more special role in stabilizing the compact folded structures of RNA pseudoknots [54±57]. However, the above structure prediction models seldom consider the conditions departing from the high salt (e.g., 1M NaCl). Although all-atomic molecular dynamics simulations can be used to probe ion-RNA interactions, it is still difficult to simulate RNA structure folding at present due to the huge computation cost [56,57]. In simplified CG models, the effect of ions (especially $Mg^{2+}$) is seldom properly involved due to the interplay between ion binding and structure deformation as well as the particularly efficient role of $Mg^{2+}$ beyond mean-field description [51±53]. Recently, a Gō-like CG model has been introduced to reproduce the folding thermodynamics of several RNA pseudoknots in the presence of monovalent ions [46,58,59], and another structure-based model can well capture the ion atmosphere around RNAs with an explicit description of divalent ions [60]. However, the two structure-based models could not be used to predict 3D structures for RNA pseudoknots solely from the sequences [11,20,46,60]. Therefore, it still remains an important problem to predict 3D structures and thermodynamic stability for RNA pseudoknots especially in monovalent/divalent ion solutions only from the sequences.

In this study, we focused on predicting 3D structures and stability for extensive RNA pseudoknots in monovalent and divalent ion solutions from their sequences through our previously developed three-bead CG model [61,62]. In the following, we first revisited the key features of our CG model such as the CG representation and the implicit-solvent/salt force field for RNAs. We then employed the model to predict 3D structures for various RNA pseudoknots from their respective sequences. Afterward, we made the prediction for the stability of typical pseudoknots with different lengths and sequences over a wide range of monovalent/divalent ion concentrations. Finally, we made the comprehensive analyses on the unfolding pathway for various RNA pseudoknots in ion solutions and examined the effect of monovalent/divalent ions on the unfolding pathway of RNA pseudoknots. Throughout the article, we have made the comparisons between the predictions and the extensive experimental data as well as the comparisons with the existing models.

## Materials and methods

### Coarse-grained structure representation for RNAs

In our model, an RNA is represented as a chain of nucleotides, where each nucleotide is reduced to three beads retaining the key structure features of an RNA chain [46,47,61,62]. As shown in Fig 1A, the backbone phosphate bead (P) and sugar bead (C) coincide with the phosphate and C4' atoms of a nucleotide, and the base beads (N) are placed at the base atoms linked to the sugar, that is N1 atom for pyrimidine or the N9 atom for purine [61,62]. The P, C and N beads are treated as spheres with van der Waals radii of 1.9Å, 1.7Å and 2.2 Å, respectively, and each P bead has a charge of $\pm e$ on its center [61,63].

### Coarse-grained force field and simulation procedure

In the CG model, the effective potential energy of an RNA conformation is given by [61,62]

$$U = U_b + U_a + U_d + U_{exc} + U_{bp} + U_{bs} + U_{cs} + U_{el}, \tag{1}$$

where bond length energy $U_b$, bond angle energy $U_a$ and dihedral energy $U_d$ account for chain connectivity and angular rotation for an RNA chain, and $U_{exc}$ represents for excluded volume interactions between two CG beads. $U_{bp}$ and $U_{bs}$ are the base-pairing and base-stacking interactions, and $U_{cs}$ is the coaxial stacking interaction between two neighbor stems. The last term $U_{el}$ corresponds to electrostatic interactions between phosphate groups, which are ignored by most of the existing predictive models for RNA 3D structures [11,20].

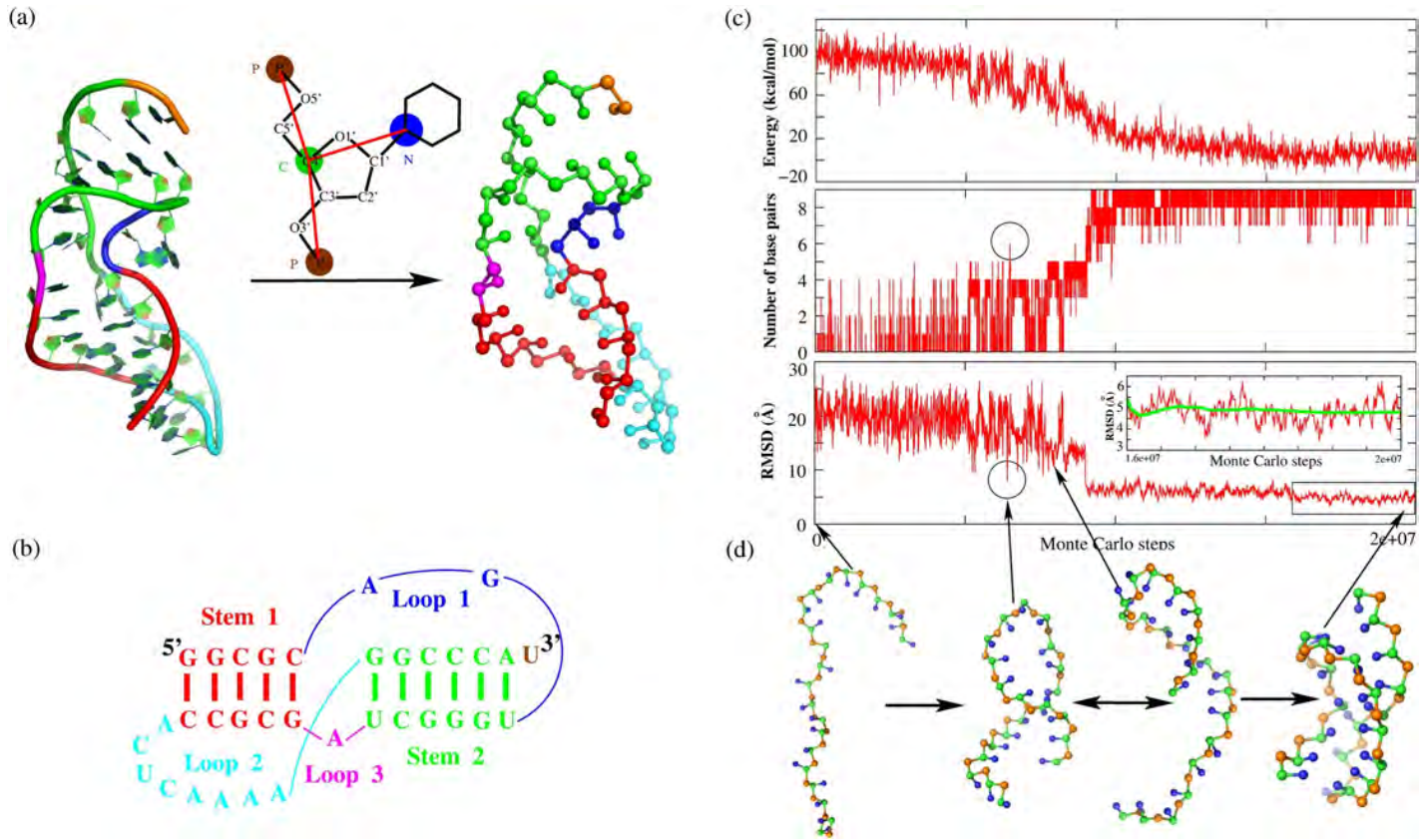

**Fig 1. The representation of all-atom and CG model for an RNA pseudoknot, and 3D structure prediction for a paradigm pseudoknot in the present model.** (a) The 3D structure of MMTV pseudoknot (PDB code: 1rnk) in all-atomistic (left) and our CG representation (right) as well as the schematic representation for one nucleotide in the present CG model (middle). (b) The secondary structure for MMTV pseudoknot consisting of two stems and three loops. The corresponding secondary structure elements in (a) and (b) are in same colors: Stem 1 (red), Stem 2 (green), Loop 1 (blue), Loop 2 (cyan), Loop 3 (magenta) and the 3' dangling nucleotide U (orange). (c, d) The time-evolution of the energy (top panel in (c)), the number of base pairs (middle panel in (c)), the RMSDs between predicted structures and the native structure in PDB (bottom panel in (c)), and the typical 3D structures (d) during the Monte Carlo simulated annealing simulation of the *Aquifex aeolicus* tmRNA pseudoknot PK1 (PDB code: 2g1w). The inset in bottom panel in (c) shows the zoomed portion of the figure in the interval of $[1.6 \times 10^7, 2 \times 10^7]$. The RMSDs are calculated over C beads from the corresponding C4' atoms in native structure, and the structures in (d) are shown with the PyMol (http://www.pymol.org).

The detailed description of the potentials in Eq 1 and the determination of the potential parameters have been described in S1 Text and also in Refs. [61,62]. Briefly, two sets of parameters of the bonded potentials ($U_b$, $U_a$ and $U_d$), Para$_{nonhelical}$ used in RNA folding process and Para$_{helical}$ used only in structure refinement for helical stems, are derived respectively from single strands/loops and stems in the PDB [12,61,64]. The sequence-dependent strength of base-staking energy is derived from the combination of the experimental thermodynamic parameters [65±67]. In most occurring pseudoknots with interhelix loop length ≤ 1nt, two helical stems can be often coaxially stacked to form a quasi-continuous double helix (Fig 1), and the strength of $U_{cs}$ depends on sequences of two interfaced base pairs [65]. The coaxial stacking could stimulate high levels of -1 frameshifting [3,4], and consequently, could be important for stabilizing functional structures of RNA pseudoknots. The electrostatic interaction $U_{el}$ is taken into account through the combination of the Debye-Hückel approximation and the concept of counterion condensation (CC) [68]. Notably, based on the tightly bound ion (TBI) model [51,52,69], the competition between monovalent and divalent ions was also taken into account in $U_{el}$ to enable the CG model to simulate RNA pseudoknot folding in mixed monovalent/ divalent ion solutions [61,62]. Although the present model has been described by us in Refs. 61 and 62, the model is still not employed for 3D structure predictions of extensive RNA pseudo-knots and it has never been used to predict the stability of RNA pseudoknots in ion solutions, especially in the presence of divalent ions [61,62]. Here, the model will be tested by extensive RNA pseudoknots on 3D structure prediction, and be further used to predict thermodynamic stability and the unfolding pathway for various RNA pseudoknots over the wide range of monovalent/divalent ion conditions.

Based on the CG force field, the Monte Carlo (MC) simulations with simulated annealing algorithm are used to predict 3D structures of RNA pseudoknot [43,61,62], where an initial simulation is started at a high temperature and a given solution condition from a totally random chain configuration generated from an RNA sequence. The system is then gradually cooled in steps, and the ion condition is fixed during the cooling process. At each temperature, RNA conformational changes are accomplished via the pivot moves which have been demonstrated to be rather efficient in sampling conformations of polymers [63], and the changes accepted or rejected according to the standard Metropolis algorithm [43,61]. The final structures obtained at the lowest target temperature (e.g., room/body temperature) are the folded conformations of the RNA predicted by the CG model. Notably, the recorded trajectories at different temperatures during the cooling process allow us to analyze the stability of the RNAs [61,62].

## Results

In this section, first, we employed the present CG model to predict the 3D structures for extensive RNA pseudoknots. Afterwards, the CG model was used to predict the stability of various RNA pseudoknots and the effects of monovalent and divalent ions. Finally, we made the comprehensive analyses on the unfolding pathway of RNA pseudoknots and the ion effect. Our predictions were compared with the available experimental data and existing models.

### Predicting 3D structures of RNA pseudoknots

**Folding process and structure refinement.** For each RNA pseudoknot, a random chain is generated from its sequence only based on the potentials of $U_b$ and $U_{exc}$ in Eq 1. Afterwards, for the random configuration, the MC simulation with simulated annealing algorithm is performed from high temperature to the target temperature (e.g., 298 K) with the use of Para$_{nonhelical}$ parameters. As an example, Fig 1C shows the folding process of a small RNA pseudoknot (22nt; PDB code: 2g1w; sequence: 5'-GGGGUGGCUCCCCUAACAGCCG-3') in the present model. As temperature is gradually decreased from 130ˆC to 25ˆC, the energy of the RNA chain reduces with the formation of base pairs (Fig 1C), and the initial random chain folds into its native-like pseudoknotted structures; see Fig 1D. Following that, another MC simulation (e.g., final $5\times10^6$ steps in Fig 1C) is performed at target temperature based on the final structure predicted by the preceding annealing process, and the two sets of bonded potential parameters Para$_{nonhelical}$ and Para$_{helical}$ are employed respectively for the single strands/loops and base-pairing regions to better capture the geometry of helical part [61]. As a result, an ensemble of refined 3D structures are obtained, and can be evaluated by their root-mean-square deviation (RMSD) values calculated over C beads from the corresponding C4' atoms in the native structure in PDB [70]; see Fig 1C. The mean RMSD (the averaged value over the structure ensemble in the refinement process) and minimum RMSD (corresponding to the structure closest to the native one in the refinement process) are used to evaluate the reliability of our predictions on 3D structures. As shown in the inset of the bottom panel of Fig 1C, the mean and minimum RMSDs of the paradigm RNA pseudoknot (PDB code: 2g1w) between predicted structures and its native structure are 4.8 Å and 3.3 Å, respectively, and the corresponding predicted 3D structures as well as the native one are also shown in Fig 2A.

**Structure prediction and comparisons with previous models.** To examine the ability of the model on predicting 3D structures of RNA pseudoknots, 17 common pseudoknots ($\leq$ 56nt) which have been determined by experiments as individual molecules were used in our 3D structure prediction. The detailed descriptions of these pseudoknots are listed in Table B in S1 Text.

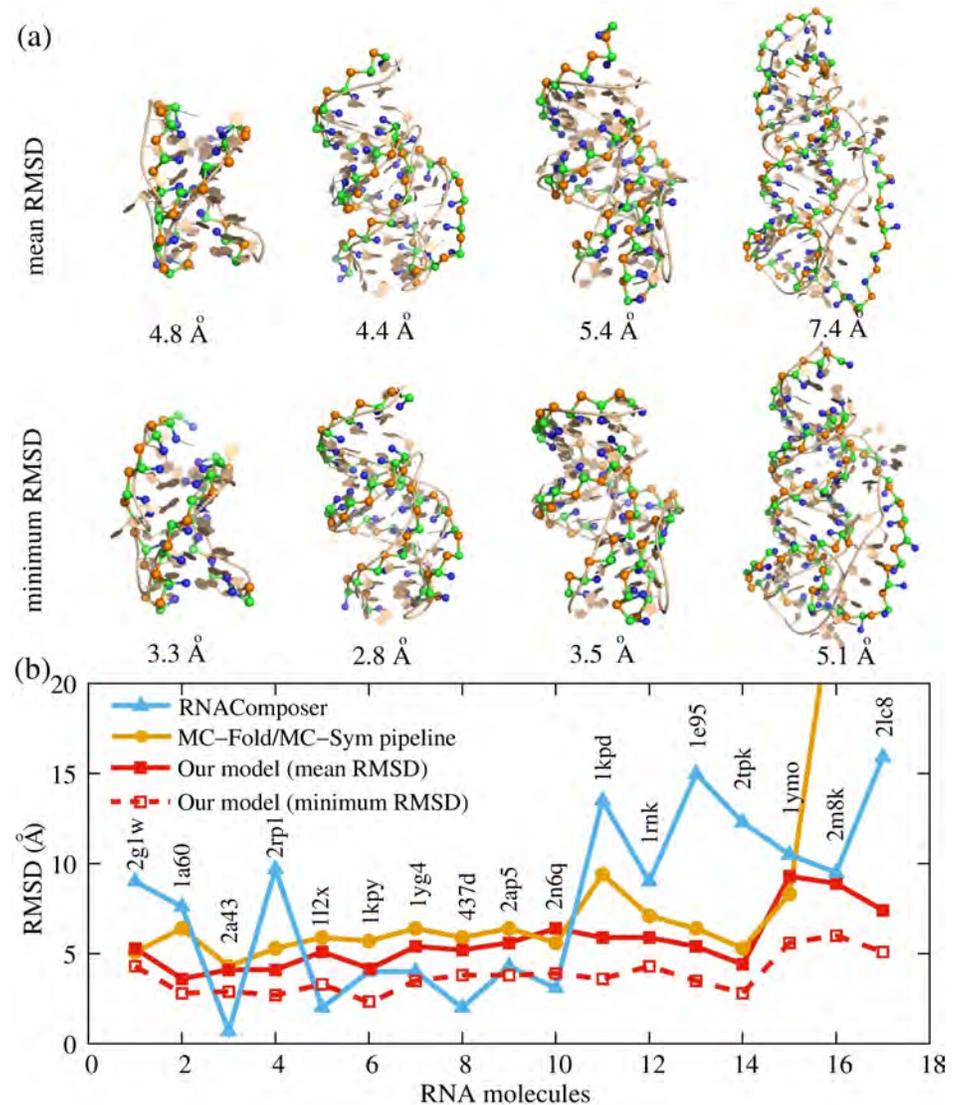

**Fig 2. Comparisons of RMSDs between the present model and other models.** (a) The predicted 3D structures (ball-stick) with the mean RMSDs (top) and the minimum RMSDs (bottom) for four sample RNA pseudoknots (PDB codes: 2g1w, 2tpk, 1e95, and 2lc8) from their native structures (cartoon). The mean (minimum) RMSDs for three pseudoknots are 4.8 Å (3.3 Å), 4.4 Å (2.8 Å), 5.4 Å (3.5 Å) and 7.4 Å (5.1 Å), respectively, and the 3D structures are shown with the PyMol (http://www.pymol.org). (b) The predictions for the 3D structures of 17 RNA pseudoknots from the present model, from the MC-Fold/MC-Sym pipeline and from the RNAComposer. The RMSDs of predicted structures for 17 RNA pseudoknots are calculated over C beads from the corresponding C4' atoms in native structures.

https://doi.org/10.1371/journal.pcbi.1006222.g002

Due to lack of the ion conditions for the experimental structures determined by X-ray crystallography, here we only predicted the 3D structures for all RNA pseudoknots at 1M [Na⁺].

Fig 2A shows the predicted 3D structures (ball-stick) with the mean and minimum RMSDs and the experimental structures (cartoon) for four typical RNA pseudoknots with different lengths and sequences. As shown in Fig 2, the present model can effectively capture the 3D shapes of RNA pseudoknots, in which Loop 1 crosses the deep major groove of the lower Stem 2, while Loop 2 generally crosses the minor groove side of Stem 1 [2,3]. The mean and minimum RMSDs for the 17 pseudoknots are shown in Fig 2B, and for most of pseudoknots, the mean and minimum RMSDs are less than 6 Å and 4 Å, respectively, which suggest that the model can make reliable predictions for 3D structures of RNA pseudoknots.

Furthermore, we also made comparisons with the MC-Fold/MC-Sym pipeline [24] and RNAComposer [30], which are well-established web services (http://www.major.iric.ca/MC-Fold/ and http://rnacomposer.ibch.poznan.pl/, respectively) for predicting tertiary structures of RNAs including pseudoknots with high accuracy [10,13]. For the 17 pseudoknots used here (Table B in S1 Text), we first employed the MC-Fold (option: consider H-type pseudoknots, return the best 1000 structures, and explore the best 50% sub-optimal structures) to predict their secondary structures in MC-Sym format, the best ones of which are further submitted (or edited and then submitted) to MC-Sym (option: model_limit = 9999 and time_limit = none) for tertiary structure predictions. The RMSDs of best structures (top 1) predicted by the MC-Fold/MC-Sym pipeline are calculated over C4′ atoms from the corresponding atoms in the experimental structures in PDB; see Fig 2B. It should be noted that although we chose the best option of MC-Fold/MC-Sym, the pipeline still fails to predict the 3D structures for two RNA pseudoknots (PDB codes: 2m8k and 2lc8), even though the experimental secondary structures are taken as input. For the 15 pseudoknots except for 2m8k and 2lc8, the overall mean RMSD from the present model is ~5.4 Å, which is smaller than that of 6.3 Å from the MC-Fold/MC-Sym pipeline, suggesting that the present model gives slightly better predictions for tested sequences. Notably, for 2m8k of 48nt and 2lc8 of 56nt, the present model also gives good predictions with mean RMSDs of 8.9 Å and 7.4Å and minimum RMSDs of 6.0 Å and 5.1 Å, respectively. Second, we further employed the RNAComposer (in interactive mode) to predict tertiary structures for the 17 pseudoknots by entering their sequences and experimental secondary structures, and the RMSD between the predicted and experimental structures is also calculated over C4′ atoms. As shown in Fig 2B, the average prediction accuracy (overall mean RMSD = ~5.6 Å and overall minimum RMSD = ~3.9 Å) of the present model for the 17 pseudoknots is slightly better than that of the RNAComposer (mean RMSD = 7.7 Å). Therefore, the comparisons with the MC-Fold/MC-Sym pipeline and RNAComposer show that the present model can make reliable predictions for 3D structures of RNA pseudoknots from sequences.

To clarify the contributions of the coaxial stacking potential, we further made the additional predictions on 3D structures for 17 RNA pseudoknots using the present model without involving the coaxial stacking potential. As shown in Table C in S1 Text, for the RNA pseudoknots except for the five ones (PDB codes: 2a43, 1l2x, 1yg4, 437d, 2ap5), the present model with coaxial stacking can make the better predictions with lower RMSDs compared to those without involving the coaxial stacking potential, which suggests that the inclusion of the coaxial stacking in the model can generally improve the 3D structure prediction for RNA pseudoknots. However, for the RNA pseudoknots such as 2a43 and 1l2x from plant luteovirus which do not have coaxial stacking interactions [4,71±74], the involvement of the coaxial stacking potential gives slightly worse predictions than those without the coaxial stacking. This suggests that the coaxial stacking potential may need to be further developed in more details for various RNA pseudoknots.

## Predicting stability of RNA pseudoknots in monovalent/divalent ion solutions

Beyond 3D structure prediction, the present model was also employed to predict the stability of RNA pseudoknots in monovalent and divalent salt solutions.

**Predicting RNA pseudoknot stability.** First, we used the MMTV frameshifting pseudoknot [75] as an example to show how to examine RNA pseudoknot stability with the use of our CG model. MMTV pseudoknot is an H-type RNA pseudoknot containing an unpaired adenosine at the junction of the two helical stems; see Fig 1B. The stability of MMTV pseudoknot directly affects the efficiency of frameshifting activity [75]. Beyond the 3D structure predictions for MMTV pseudoknot (PDB code: 1rnk) (Fig 2), we further employed the present model to examine the stability of MMTV pseudoknot. Fig 3A shows that the number of formed base pairs changes at different temperatures, and there are mainly three states over temperature: the fully folded pseudoknot state (F) at low temperatures (e.g., <~40ÊC);the unfolded coil state (U) at high temperatures (e.g., >~100ÊC);and partially unfolded intermediate hairpin states (I) at medium temperature which is coincident with experiments [75±77]. Then, the fractions of the three states at each temperature can be calculated (Fig 3B), and the fractions of folded and unfolded states ($f_F(T)$ and $f_U(T)$) can be fitted to a two-state model through the following equations [61,65]:

$$f_F(T) = \frac{1}{1 + e^{(T - T_{m1})/dT_1}};$$ (2)

$$f_U(T) = 1 - \frac{1}{1 + e^{(T - T_{m2})/dT_2}}.$$ (3)

Here, $T_{m1}$ and $T_{m2}$ are two melting temperatures of the corresponding transitions (F→I and I→U), respectively. $dT_1$ and $dT_2$ are corresponding adjustable parameters. As shown in Fig 3, for MMTV pseudoknot at 50mM [K$^+$], the predicted $T_{m1}$ and $T_{m2}$ are 49.0ÊC and 81.9ÊC, which agree well with the experimental data (50.2ÊC and 83.0ÊC) [75].

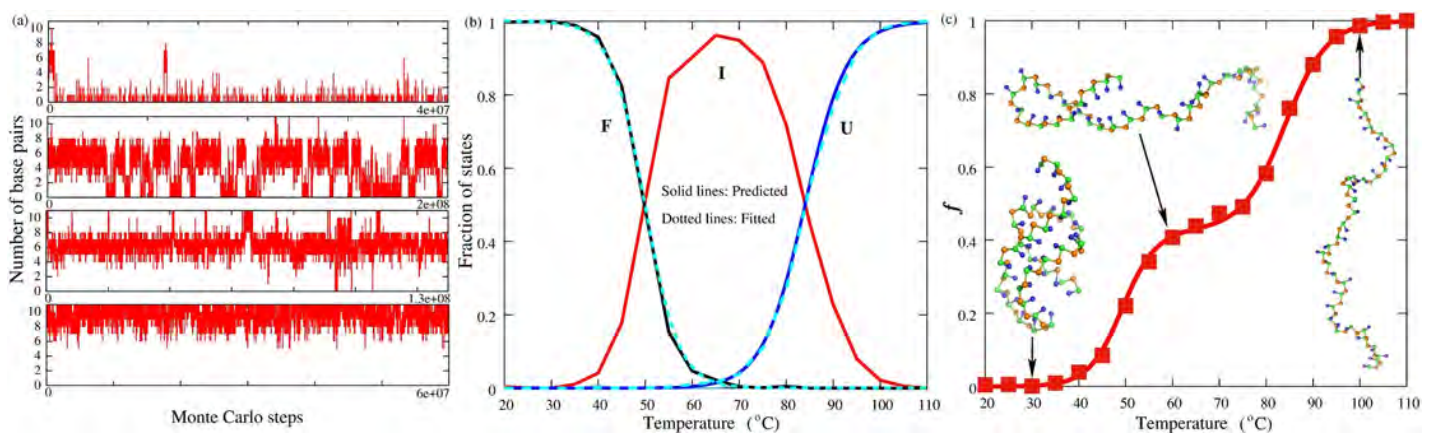

**Fig 3. The stability prediction for a sample RNA pseudoknot in the present model.** (a) The time-evolution of the number of base pairs for MMTV pseudoknot (shown in Fig 1A) at different temperatures (100ÊC,80ÊC,60ÊC,40ÊC from top to bottom, respectively) in 50mM KCl solution. (b) The fractions of folded state (F, black), unfolded state (U, blue), and intermediate state (I, red) as a function of temperature for MMTV pseudoknot at 50mM [K$^+$]. The dotted lines are fitted to the predicted data through Eqs. 2 and 3. (c) The fraction of denatured base pairs $f$ as a function of temperature for MMTV pseudoknot at 50mM [K$^+$]; symbols: from the present model; line: from Eq 4. Ball-stick: the typical 3D structures predicted at different temperatures shown with the PyMol (http://www.pymol.org).

https://doi.org/10.1371/journal.pcbi.1006222.g003

Furthermore, based on the $f_F(T)$ and $f_U(T)$, the fraction of number of denatured base pairs $f$ can be calculated through the following equation [76]

$$f = 1 - [(1 - f_I) \cdot f_F(T) + f_I \cdot (1 - f_U(T))],\qquad(4)$$

where $f_I$ is the fraction of number of denatured base pairs when the fraction of intermediate state is maximum; see Fig 3C. As shown in Figs 4A and 5A, $df/dT$ (the first derivative of $f$ calculated by Eq 4 with respect to temperature) profile for MMTV pseudoknot is in good agreement with previous differential scanning calorimetry profile [75]. This suggests that the present model can give reliable predictions for stability of the RNA pseudoknot.

In addition to MMTV pseudoknot, five other pseudoknots with different sequences from gene 32 mRNA [77] and plant luteoviruses [71±74] are also examined by the present model, and the sequences and predicted secondary structures for the pseudoknots can be found in S1 Fig. As shown in Table 1, our predictions on melting temperatures ($T_{m1}$ and $T_{m2}$) for the six RNA pseudoknots at high salt concentrations (1000mM [K+] for MMTV and T2, and 500mM [K+] for the other four pseudoknots) agree well with the experimental data with the mean deviations of ~2.5ÊCfor $T_{m1}$ and ~1.8ÊCfor $T_{m2}$. In addition, Fig 4 shows the comparisons between the calculated and the experimental thermal unfolding curves [71±77] for the six RNA pseudoknots at high salt, indicating that the predicted melting curves are also in good accordance with the experiments over different sequences. These suggest that the present model with stacking interactions parametrized using a set of sequence-specific thermodynamics parameters can quantitatively predict the stability of RNA pseudoknots with various sequences. Nevertheless, the predicted $T_{m1}$'s for BWYV and PEMV-1 are slightly lower than

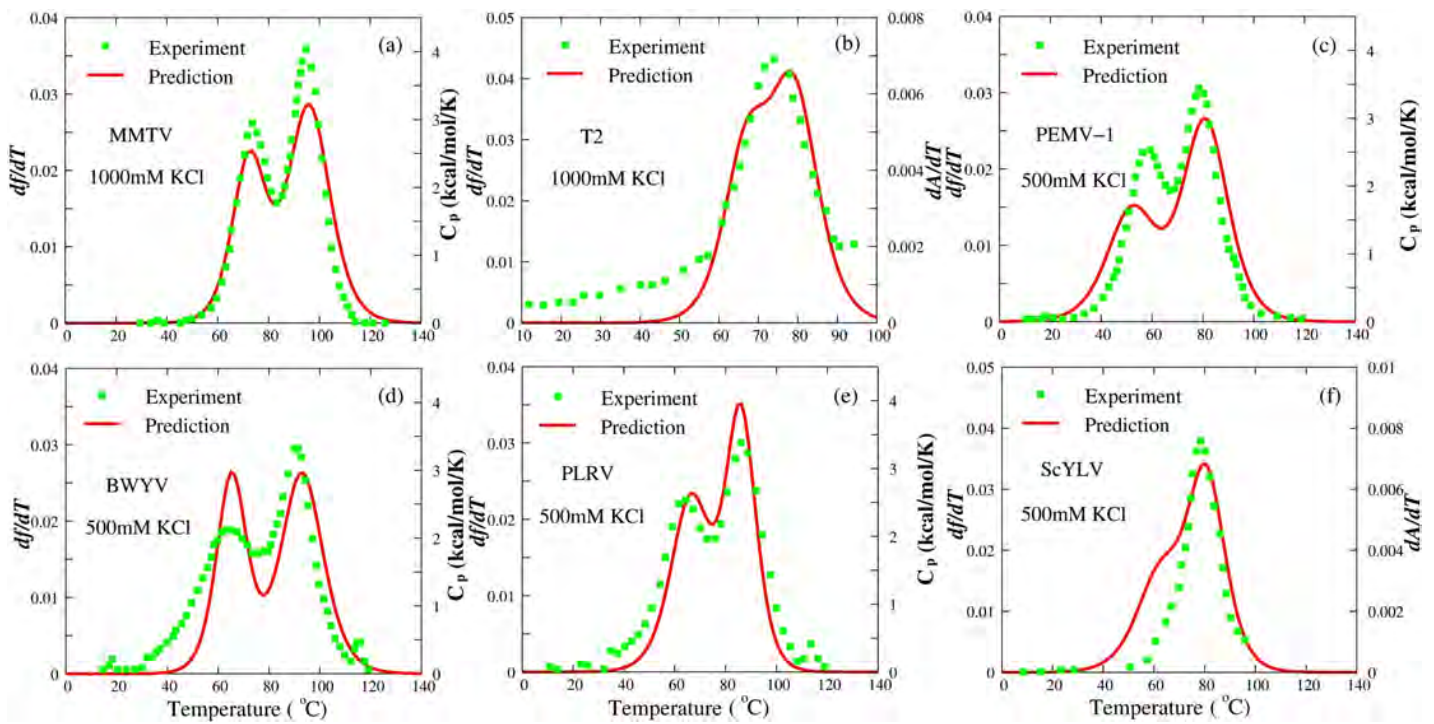

Fig 4. The comparisons between predictions (lines) and experiments (symbols) for six RNA pseudoknots with various sequences at high salt concentrations. (a) MMTV pseudoknot at 1000mM [K+] [75]; (b) T2 pseudoknot at 1000mM [K+] [77]; (c) PEMV-1 pseudoknot at 500mM [K+] [73]; (d) BWYV pseudoknot at 500mM [K+] [54,72]; (e) PLRV pseudoknot at 500mM [K+] [73]; and (f) ScYLV pseudoknot at 500mM [K+] [74]. Lines: $df/dT$, the first derivative of $f$ with the temperature. Symbols: the heat capacity $C_p$, or $dA/dT$, the first derivative of absorbance with respect to temperature.

https://doi.org/10.1371/journal.pcbi.1006222.g004

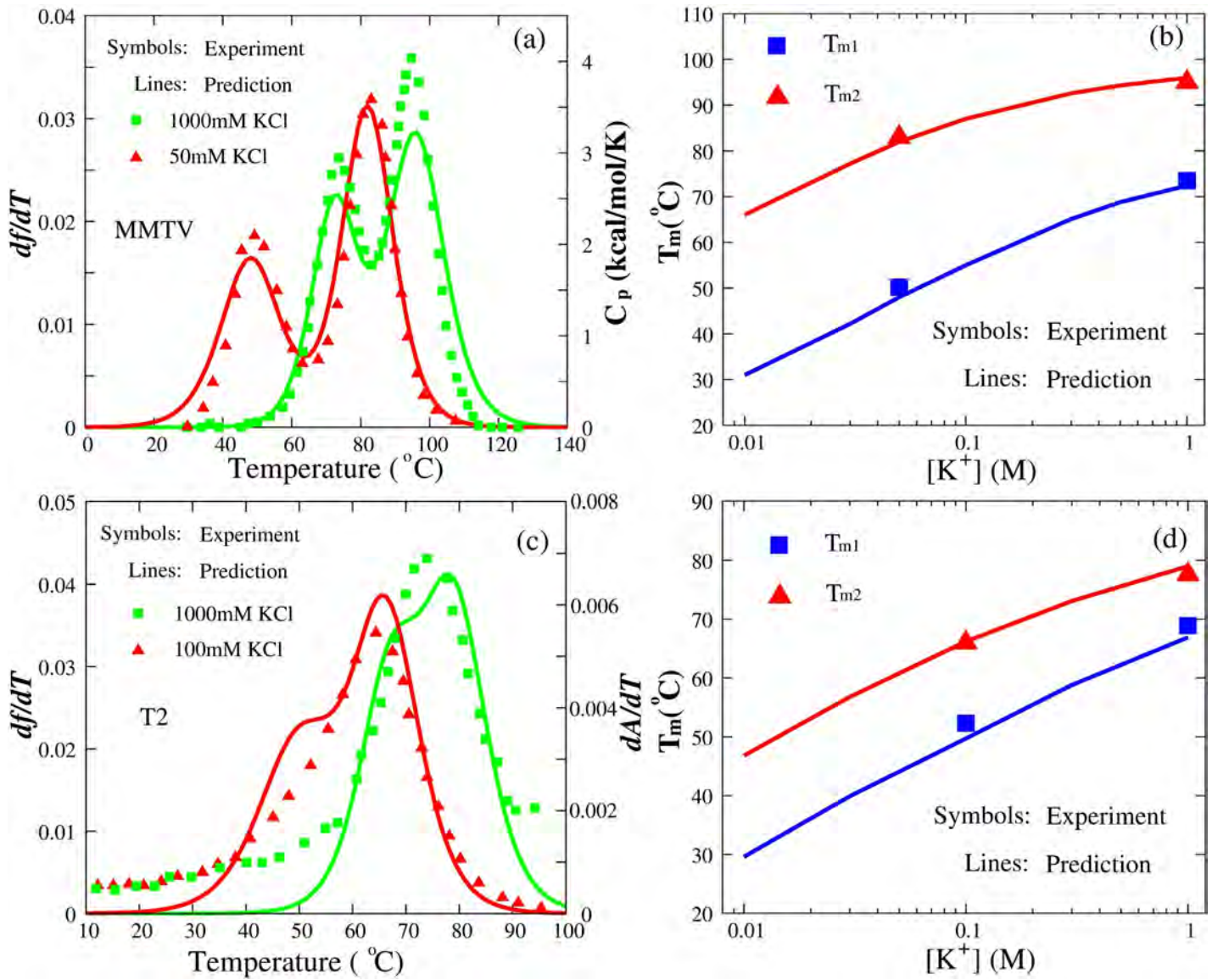

**Fig 5. The comparisons between predictions (lines) and experiments (symbols) for MMTV and T2 pseudoknots in monovalent ion solutions.** (a) MMTV pseudoknot at 1000mM [K$^+$] (green square) and 50mM [K$^+$] (red triangle), respectively. Lines: $df/dT$, the first derivative of $f$ with respect to temperature. Symbols: the heat capacity $C_p$. (c) T2 pseudoknot at 1000mM [K$^+$] (green square) and 100mM [K$^+$] (red triangle), respectively. Symbols: $dA/dT$, the first derivative of absorbance with respect to temperature. (b, d) The melting temperatures $T_{m1}$ and $T_{m2}$ of two transitions (F→I and I→U) as functions of [K$^+$] for MMTV (b) and T2 (d) pseudoknots. Symbols: experimental $T_{m1}$ (blue square) and $T_{m2}$ (red triangle) [75,77]. Lines: corresponding predictions.

the corresponding experimental data; see Table 1. One possible reason is that both pseudoknots contain a collection of loop-stem tertiary structural interactions, which could make contributions to the stability of the pseudoknotted structures [4,72–74].

To further evaluate the present model, we made the comparisons with the available predictions from the existing models such as the model of Denesyuk and Thirumalai [46,58,59] and the Vfold model with monovalent salt-corrected thermodynamic parameters and the fitting parameters of loop-stem tertiary contacts [37,38,78], for the stability of MMTV, BWYV and PEMV-1 pseudoknots in monovalent ion solutions. As shown in S2 Fig, for BWYV and PEMV-1 pseudoknots at 500mM [K$^+$], the mean deviation between $T_m$'s ($T_{m1}+T_{m2}$) from the

**Table 1. The melting temperatures ($T_{m1}$ and $T_{m2}$)[a] of six RNA pseudoknots at high salt concentrations.**

| RNA pseudoknots[b] | References | Expt. (°C) $T_{m1}/T_{m2}$ | Pred. (°C) $T_{m1}/T_{m2}$ | Deviation (°C) $|\Delta T_{m1}|/|\Delta T_{m2}|$ |
|---|---|---|---|---|
| MMTV | 75 | 73.5/95.0 | 71.7/96.3 | 1.8/1.3 |
| T2 | 77 | 68.9/77.6 | 67.2/78.8 | 1.7/1.2 |
| PEMV-1 | 73 | 60.1/79.1 | 54.5/80.8 | 5.6/1.7 |
| BWYV | 54,72 | 69.4/91.2 | 65.8/93.1 | 3.6/1.9 |
| PLRV | 73 | 67.4/87.5 | 66.3/85.3 | 1.1/2.2 |
| ScYLV | 74 | 67.5/77.9 | 66.2/80.2 | 1.3/2.3 |

[a] $T_{m1}$ and $T_{m2}$ are the melting temperatures for the transitions from folded state to intermediate state and from intermediate state to unfolded state, respectively.
[b] MMTV and T2 pseudoknots at 1000mM [$K^+$]; PEMV-1, BWYV, PLRV and ScYLV pseudoknots at 500mM [$K^+$].

https://doi.org/10.1371/journal.pcbi.1006222.t001

Vfold model and the experiments is ~8.4°C[78], which is slightly higher than that from the present model (~6.4°C).For MMTV pseudoknot at 1000mM [$K^+$], the deviation between $T_m$'s ($T_{m1}+T_{m2}$) from the model of Denesyuk and Thirumalai and the experiment is ~8.2°C[46], a higher deviation than that from the present model (~3.1°C).Such deviation from the model of Denesyuk and Thirumalai and the present model becomes rather small for MMTV at 50mM [$K^+$]. Thus, the present model can be reliable in predicting thermal stability for RNA pseudoknots in monovalent ion solutions, and it is noted that the present model can also provide 3D structures for RNA pseudoknots at different temperatures from the sequences; see Fig 3C and S3 Fig. Recently, the HiRE-RNA model and the oxRNA model have been proposed and both can predict the presence of two peaks in the melting curves of several RNA pseudoknots, however, there is still lack of the quantitative comparisons between the two models and the experiments for extensive RNA pseudoknots [40,42].

Furthermore, we made the additional calculations for the stability of the six RNA pseudoknots (Table 1 and Table D in S1 Text) using the present model without involving the coaxial stacking potential. As shown in Table 1, S3 Fig and Table D in S1 Text, $T_{m1}$'s from the present model with coaxial stacking are generally higher than those without coaxial stacking and appear closer to the experimental values [72÷77], suggesting that the involvement of coaxial stacking enhances the stability of RNA pseudoknotted structures and improves our predictions on RNA pseudoknot stability. It is also noted that, in some pseudoknots such as BWYV pseudoknot, no coaxial stacking is found between two stems, while triple/quadruple base interactions are observed at the junctions in their experimental structures [71÷74]. It is interesting that the present model without considering triple/quadruple interactions still make good predictions on the stability for this kind of pseudoknots. This should be attributed to that the involvement of coaxial stacking in the present model partially compensates the lack of triple/quadruple base interactions in the model.

**Monovalent ion effect on RNA pseudoknot stability.** Due to the high density of negative charges on its backbone, RNA pseudoknot stability is sensitive to ionic condition of solution [49,51÷54,77,79], while the effect of salts is generally ignored in the existing RNA structure prediction models [11,20]. For MMTV pseudoknot, we further performed the simulations at different temperatures over a broad range of [$K^+$], and calculated the melting temperatures based on the data from the simulations. As shown in Fig 5A, the unfolding of MMTV pseudoknot at different [$K^+$]'s has the similar two transitions in accordance with the available experiments [75], and the predicted melting temperatures ($T_{m1}$ and $T_{m2}$) agree well with the experimental data with the maximum deviation of ~1.8°Cover different $K^+$ concentrations; see Fig 5B.

In addition, we examined the stability of the bacteriophage T2 gene 32 mRNA pseudoknot in solutions of different [K$^+$]'s. The sequence and predicted secondary structure of T2 pseudoknot are shown in S1 Fig, and the predicted and experimental 3D structures (PDB code: 2tpk) are shown in Fig 2. As shown in Fig 5C, in contrast to MMTV pseudoknot, there is one visible peak in experimental unfolding curves for T2 pseudoknot at 100mM [K$^+$] and 1000mM [K$^+$] due to the smaller difference between two transition temperatures $T_{m1}$ and $T_{m2}$ for T2 pseudoknot [75,77], i.e, at 1M [K$^+$], $\Delta = |T_{m1}$-$T_{m2}| \approx 12$°C for T2 pseudoknot and ~24°C for MMTV pseudoknot. In addition, the predicted $T_{m1}$ and $T_{m2}$ for T2 pseudoknot are also in good accordance with the experimental data with the mean deviation of ~1.1°C over the wide range of [K$^+$] [77]; see Fig 5D.

As shown in Fig 5B and 5D, the increase of [K$^+$] enhances the stability of RNA pseudoknots as well as the stability of intermediate hairpins, and $T_{m1}$ of the transition from folded state to intermediate state is more sensitive to [K$^+$] than $T_{m2}$ of the transition from intermediate state to fully unfolded chain. The phenomenon is interesting and is reasonable. Generally, the formation of a pseudoknot structure involves higher charge buildup than the formation of intermediate hairpin states, and hence, the stability of a pseudoknot is more dependent on ions than that of a hairpin [51–54].

**Divalent ion effect on RNA pseudoknot stability.** Previous studies have shown that divalent ions such as Mg$^{2+}$ are especially effective in stabilizing RNA tertiary structure [48,52–54,79]. As described in Section of Material and methods, the effect of divalent ions has been implicitly accounted for in the present model by combining the CC theory [68] and the results from the TBI model [51,69]. Here, beyond previous computational models for RNA pseudoknot stability in monovalent ion solutions, the present model is examined by predicting the stability for two typical RNA pseudoknots (MMTV and T2) in mixed monovalent/divalent ion solutions. As shown in Fig 6, the thermal unfolding curves and $T_m$'s predicted by the present model for MMTV and T2 pseudoknots are in accordance with the experiments [75,77] over a wide range of [Mg$^{2+}$]'s with fixed [K$^+$]'s (50mM for MMTV pseudoknot and 100mM for T2 pseudoknot). For MMTV pseudoknot, the mean deviations of $T_{m1}$ and $T_{m2}$ between predictions and experiments are ~3.9°C and ~2.3°C, respectively, and for T2 pseudoknot, the corresponding mean deviations are ~2.2°C for $T_{m1}$ and ~1.5°C for $T_{m2}$. This suggests that the present model can nearly make quantitative predictions for the stability of RNA pseudoknots in mixed ion solutions from their sequences, even though the ion effect is involved implicitly in the present model.

Fig 6 also shows that the competition between K$^+$ and Mg$^{2+}$ on RNA pseudoknot stability is captured by the present model: (i) when [Mg$^{2+}$] is very low, K$^+$ ions dominate the stability of pseudoknots and the values of $T_m$ including $T_{m1}$ and $T_{m2}$ are close to those of pure K$^+$ solutions; (ii) the increase of [Mg$^{2+}$] significantly enhances the stability of RNA pseudoknots against monovalent K$^+$ and such effect would become saturated at very high [Mg$^{2+}$] due to the strong electrostatic neutralization; see Fig 6B and 6D. This is attributed to the anticooperative binding of K$^+$ and Mg$^{2+}$ and more efficient role of Mg$^{2+}$ binding [49–52]. Nonetheless, the predicted melting temperatures, e.g., $T_{m1}$ of the first transition, are slightly lower than the corresponding experimental values at high [Mg$^{2+}$]; see Fig 6B and 6D. This may be attributed to the fact that the implicit Mg$^{2+}$ treatment in the present model might slightly underestimate the role of Mg$^{2+}$ in stabilizing compact pseudoknot structure [53].

## Thermally unfolding pathway of RNA pseudoknots in ion solutions

Since intermediate states of RNAs can be important to their biological functions [5,76,80–85], unfolding pathway of RNAs including some pseudoknots has been studied through theoretical modeling and experiments [75±77,81±88]. To examine the unfolding pathway of RNA pseudo-knots, we made comprehensive analyses for six RNA pseudoknots; see Fig 7 and S4 Fig. Based on the simulations for each pseudoknot at a given solution condition, beyond the fractions of states F and U, the fractions of different intermediate hairpin states (named as S1 and S2 for intermediate states reserving one of Stem 1 and Stem 2, respectively) at different temperatures can also be calculated; see Figs 7 and 8 and S4 and S6 Figs. Furthermore, we employed the model to predict the unfolding pathway for various RNA pseudoknots in monovalent/divalent ion solutions and examined the effect of monovalent/divalent ions on the unfolding pathway of RNA pseudoknots, which was seldom covered in previous studies since the effect of divalent ions is generally difficult to be involved.

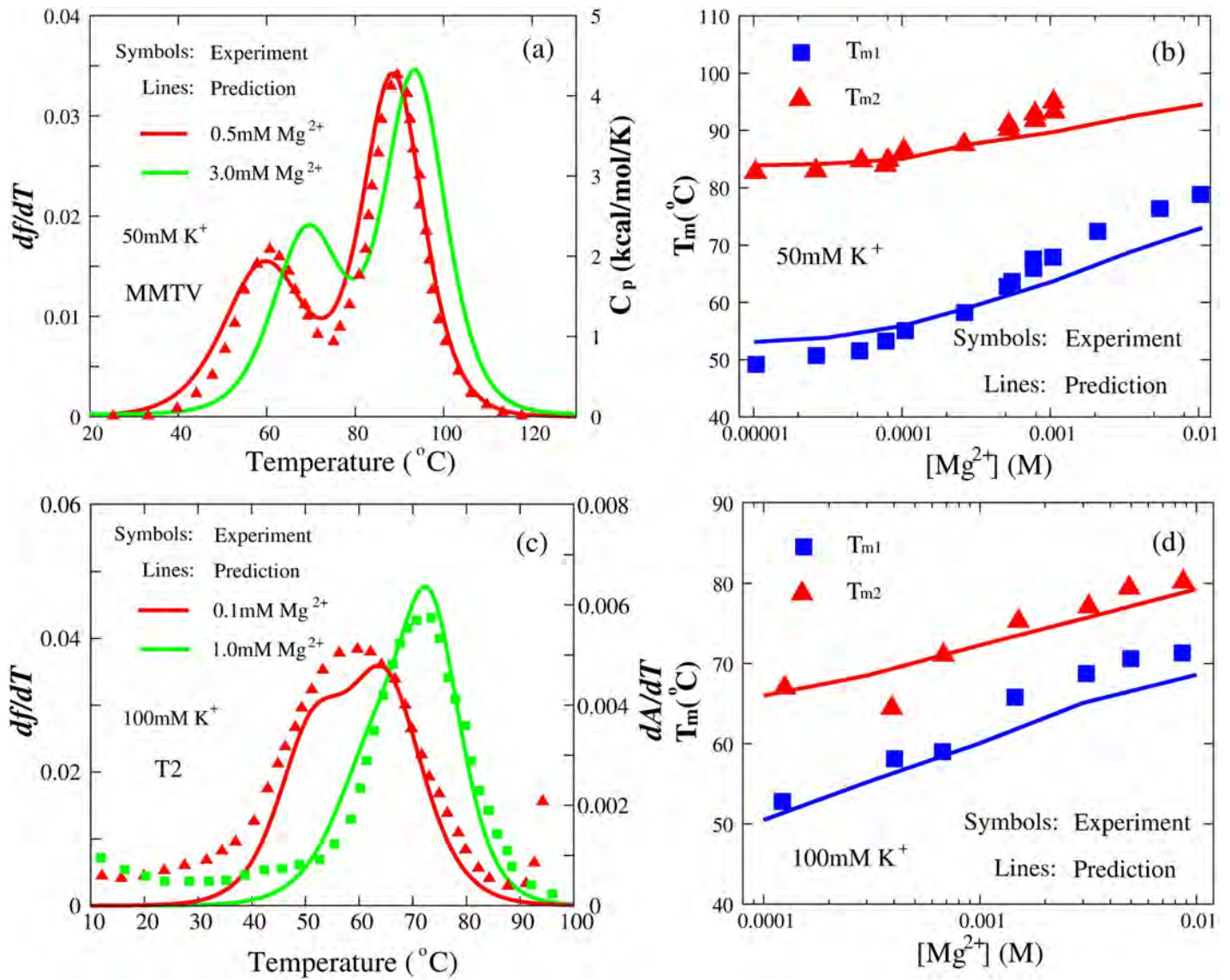

**Fig 6. The comparisons between predictions (lines) and experiments (symbols) for MMTV and T2 pseudoknots in divalent ion solutions.** (a) MMTV pseudoknot at 50mM [K⁺] with 0.5mM [Mg²⁺] and 3.0mM [Mg²⁺], respectively. Symbols: the heat capacity $C_p$ for MMTV pseudoknot at 50mM [K⁺] with 0.5mM [Mg²⁺]. (c) T2 pseudoknot at 100mM [K⁺] with 0.1mM [Mg²⁺] and 1.0mM [Mg²⁺], respectively. Symbols: $dA/dT$, the first derivative of absorbance with respect to temperature. (b, d) The melting temperatures $T_{m1}$ and $T_{m2}$ of two transitions (F→I and I→U) as functions of [Mg²⁺] for MMTV pseudoknot in the presence of 50mM [K⁺] (b) and T2 pseudoknot in the presence of 100mM [K⁺] (d). Symbols: experimental $T_{m1}$ (blue square) and $T_{m2}$ (red triangle) [75,77]. Lines: corresponding predictions.

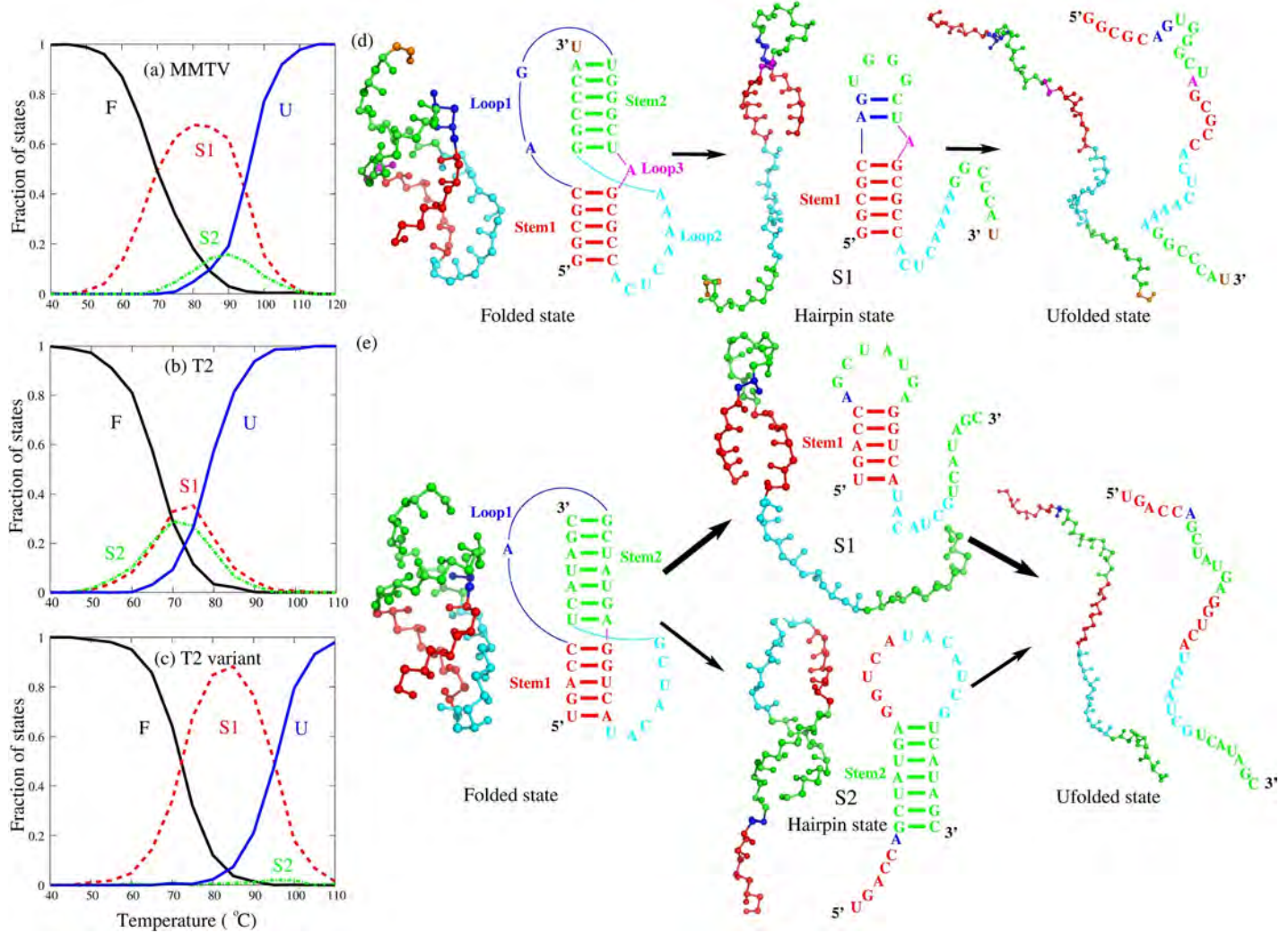

**Fig 7. The predicted thermal unfolding of MMTV, T2 and T2 variant pseudoknots at 1M [K⁺].** (a-c) The fractions of F, S1, S2 and U states as functions of temperature in thermal unfolding of MMTV (a), T2 (b), and T2 variant (c) pseudoknots at 1M [K⁺]. F stands for fully folded RNA; S1, hairpin intermediate with Stem 1; S2, hairpin intermediate with Stem 2; U, fully unfolded RNA. (d) Schematic diagram shows the dominating structural transitions of MMTV pseudoknot along the unfolding pathway inferred from the fraction of states shown in panel (a). (e) Schematic diagram shows the structural transitions of T2 pseudoknot and the variant of T2 pseudoknot along the unfolding pathway inferred from their respective fraction of states shown in panels (b) and (c).

**Unfolding pathway of RNA pseudoknot varies with its sequence.** As shown in Fig 7A for MMTV pseudoknot at 1M [K⁺], at a low temperature (e.g., <~40°C), the RNA is completely in folded pseudoknot state with two structural motifs of Stem 1 and Stem 2. As temperature is increased (e.g., 40°C-80°C), the fraction of F state decreases gradually, and simultaneously, the fraction of S1 state increases gradually, which indicates that the Stem 2 in the pseudoknot melts first with increasing temperature. When temperature is increased to higher level (e.g., >~80°C), the fraction of completely unfolded conformations increases accompanied with the decrease of the fraction of S1 state. Although the S2 state can also be found in this process, the fraction of S2 state relative to all intermediate states is relatively small (~18%). Thus, with the increase of temperature, the dominating unfolding pathway of MMTV pseudoknot is F→S1→U overwhelming the pathway of F→S2→U; see Fig 7. But for the four (PEMV-1, BWYV, PLRV and ScYLV) pseudoknots from plant luteoviruses, the unfolding processes almost undergo the only pathway F→S1→U through the intermediate state S1 and the other pathway of F→S2→U appears negligibly, as shown in S4 Fig. Interestingly, for T2 pseudoknot at 1M [K⁺], there are two comparable unfolding pathways: F→S1→U and F→S2→U, where the population of S1 state is only slightly higher than that of S2 state; see Fig 7B. Our calculations are in accordance with the experiments [75±77,80±83] as well as the recent theoretical studies [59,84±88]. The above analyses show that RNA pseudoknots of different sequences can have very different unfolding pathways.

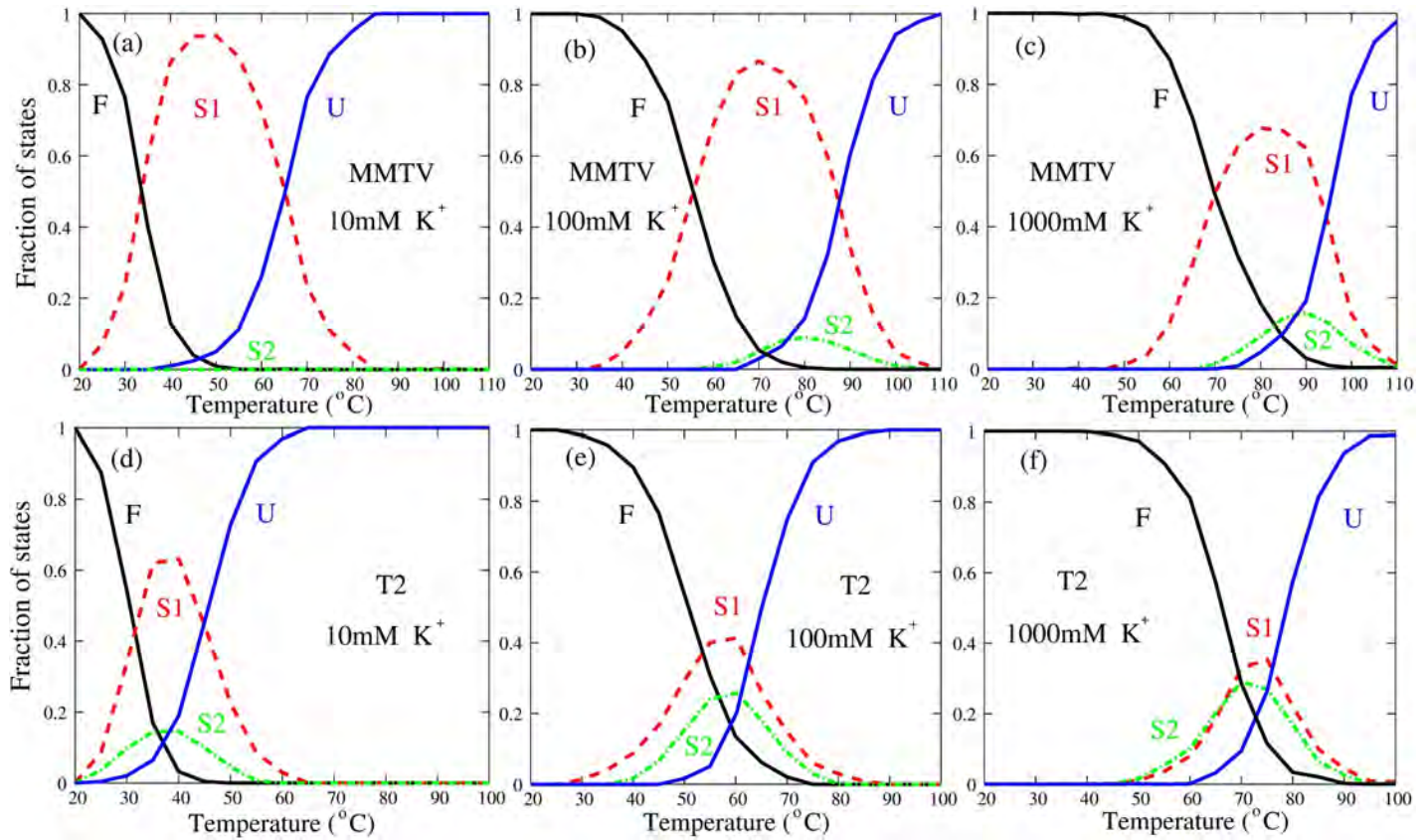

**Fig 8. The fractions of F, S1, S2 and U states as functions of temperature in thermal unfolding of MMTV (a-c) and T2 (d-f) pseudoknots at different [K⁺]'s.** F stands for fully folded RNA; S1, hairpin intermediate with Stem1; S2, hairpin intermediate with Stem2; U, fully unfolded RNA.

**What dominates the unfolding pathway of RNA pseudoknots?.**    It is interesting that different RNA pseudoknots can have apparently different unfolding pathways. Thirumalai and his coworkers have proposed that the unfolding pathway of several pseudoknots is largely determined by the stabilities of constituent secondary structures [59,84,86]. To examine that and to further understand the dominant factor on the unfolding pathway for extensive RNA pseudoknots, we calculated the free energies of the two intermediate states (hairpins) for each of these pseudoknots at different temperatures and 1M [Na⁺] using the Mfold algorithm [89]; see S5 Fig. For the pseudoknots except for T2, the two intermediate hairpin states are sufficiently different in stability, and the unstable hairpin (S2 state) containing Stem 2 with apparently higher free energy would melt first when temperature is increased. For example, the relative free energy $\Delta\Delta G = \Delta G_{S1} - \Delta G_{S2}$ between the intermediate states S1 and S2 of MMTV pseudoknot is ~-0.6kcal/mol at T~85ÊCaround which fractions of S1 and S2 have maximum values, while for the other four (PEMV-1, BWYV, PLRV and ScYLV) pseudoknots, $\Delta\Delta G$ is

~-4.0kcal/mol, ~-3.6kcal/mol, ~-2.7kcal/mol and ~-3.2kcal/mol in the range of 40±90°C, respectively. That is why the intermediate S2 state occurs with a lower and visible population in unfolding process of MMTV pseudoknot and it almost never appears in unfolding processes of other four (PEMV-1, BWYV, PLRV and ScYLV) pseudoknots. Unlike the above described pseudoknots, the two intermediate hairpins of T2 pseudoknot are very similar in stability with the relative free energy $\Delta\Delta G$ of ~0kcal/mol (in the range of 70±80°C),which would lead to two parallel unfolding pathways with similar probabilities through the intermediate states of S1 and S2, respectively. To further examine the role of stability of intermediate state on unfolding pathway, we also made the analysis for a variant of T2 pseudoknot, in which the $A_3$-$U_{16}$ base pair in Stem 1 is substituted by a more stable G-C base pair; see Fig 7C and S5 Fig. Due to the significant enhancements of stability for S1 state containing Stem 1 with $\Delta\Delta G$~-3.4kcal/mol (in the range of 70±80°C),the intermediate state of S2 nearly disappears, as shown in Fig 7C. As a result, the unfolding pathway of the variant of T2 pseudoknot is F→S1→U, which is distinctively different from that of T2 pseudoknot.

Therefore, the above comprehensive analyses for six wild-type pseudoknots and a variant of T2 pseudoknot show that the unfolding pathway of an RNA pseudoknot is mainly dependent on the stability of intermediate states, which is in consistent with the conclusion from Thirumalai and his coworkers [59,84,86]. Furthermore, despite the same stability of the two intermediate hairpins in T2 pseudoknot, the experiments show that Stem 2 can be slightly easier to melt in unfolding process [77,82,83], which indicates that other factors may affect the unfolding of pseudoknots. For example, Wang, Zhang and their coworkers have recently proposed that the contribution of the noncanonical interactions between helices and loops may make contribution to the unfolding of pseudoknots [85,88], while such interaction is not involved in the present model.

**Unfolding pathway of RNA pseudoknots can be modulated by ions.** Since ions can significantly affect the stability of intermediate states [52,53,69], we further examined the effect of monovalent/divalent ions on the unfolding pathway of MMTV and T2 pseudoknots, beyond the previous analyses on unfolding pathway of RNA pseudoknots. As shown in Fig 8 and S6 Fig, as $K^+$ concentration decreases, the structure melting transitions can get easier at low temperatures due to the enhancement of electrostatic repulsion in pseudoknots [52,53], and correspondingly, the unfolding pathway of the two pseudoknots changes due to the change of ion condition. For example, for MMTV pseudoknot, S2 state nearly never appears at low salt (e.g., 10mM $K^+$), while gradually appears as the [$K^+$] increase, i.e., the fraction of S2 state increases from ~0% at 10mM [$K^+$] to ~18.0% at 1M [$K^+$]; see Fig 8. Such prediction is in good accordance with the very recent experimental results (Roca, Hori, Velmurugu, Narayanan, Narayanan, Thirumalai, and Ansari, arXiv: 1710.0695). This is because the increase of [$K^+$] stabilizes S2 state more pronouncedly than S1 state since S2 state has a large hairpin than S1 state (excluding dangling tails), i.e., the relative free energy $\Delta\Delta G$ calculated from the salt extension from the TBI model is ~-2.5kcal/mol at 10mM [$K^+$] (at $T$~45°C)and becomes ~-0.6kcal/mol at 1M [$K^+$] (at $T$~85°C)[55,69]. For T2 pseudoknot, as $K^+$ concentration is decreased from 1M, the probabilities of the two parallel pathways change visibly, i.e., the fraction of S2 decreases from ~44.0% at 1M [$K^+$] to ~18.5% at 10mM [$K^+$]. This is also attributed to the larger hairpin of S2 state and the corresponding stronger ion effect in structure stabilization, i.e., the relative free energy $\Delta\Delta G$ is ~-0.1kcal/mol at 1M [$K^+$] (at $T$~75°C)and becomes ~-1.6kcal/mol at 10mM [$K^+$] (at $T$~40°C)[55,69]. Similarly, as shown in S6 Fig, $Mg^{2+}$ can also significantly affect the unfolding pathway of MMTV and T2 pseudoknots although they are in the buffers containing 50mM and 100mM $K^+$, respectively.

The above analyses for MMTV and T2 pseudoknots indicate that the change of ion conditions can apparently modulate the unfolding pathway of RNA pseudoknots through changing the relative stability between the two unfolding intermediate states at different ion conditions.

## Discussion

It is important to predict 3D structures and stability of RNA pseudoknots in monovalent/divalent ion solutions from their sequences. In this work, we employed our previously developed model to address this problem. Beyond mainly focusing on reproducing structures, as many previous structure prediction models have done, the present model enables us to predict and analyze 3D structure stability for RNA pseudoknots in different monovalent/divalent ion solutions. The following are the major conclusions:

1. The present model predicts the native-like 3D structures for RNA pseudoknots with an overall mean RMSD of 5.6 Å and an overall minimum RMSD of 3.9 Å from experimental structures, and the overall prediction accuracy of our model is slightly higher than previous models.

2. The present model successfully predicts the stability of RNA pseudoknots with different lengths and sequences over a wide range of monovalent/divalent ion concentrations, and the predicted melting temperatures for the two unfolding transitions are in good accordance with extensive experiment data.

3. Our comprehensive analyses show that the unfolding pathway of RNA pseudoknots is mainly determined by the stabilities of intermediate states which can be significantly modulated by the sequences and solution ion conditions.

Despite the extensive agreements between our predictions and experiments, the present model has several limitations that should be overcome in future model development. First, the present model does not treat possible noncanonical interactions such as base triple interactions between loops and stems, self-stacking in loop nucleotides and special hydrogen bonds involving phosphates and sugars, which could be important for some more complex pseudoknotted structures [7,17,38]. Beyond the common H-type pseudoknots ($\leq 56$nt) used in this work, larger RNAs with complex structures should be incorporated in to further improve the present model [19,90±94]. Second, the effect of monovalent/divalent salts is implicitly accounted for in the present model by the combination of CC theory and the TBI model. Such implicit-salt treatment may be responsible for the underestimation on the stability of RNA pseudoknots at high [Mg$^{2+}$]. Mg$^{2+}$ can play an efficient and special role in stabilizing compact RNA structures [51±54,79], and further development may need to involve Mg$^{2+}$ explicitly in our model. Third, in this work, we mainly focused on the 3D structures and thermodynamic stability of RNA pseudoknots, and did not involve the stability under mechanical force. Mechanical forces can be not only considered as a useful probe for RNA stability, but also important for the functions of some RNA pseudoknots [3,81,86,95±97]. For example, the frameshifting efficiency may be affected by the magnitude of unfolding force for RNA pseudoknots [3,81,96]. Fortunately, the present model can be extended to study the mechanical stability of RNA pseudoknots by including external force in the energy functions of the model [67,86,95]. Finally, the 3D structure predicted by the present model is at the CG level, and it is still necessary to develop the model to reconstruct all-atomistic structures based on the CG structures for further practical applications. Nevertheless, the present model could be a reliable predictive model for predicting 3D structures and stability of RNA pseudoknots in ion solutions from their sequences and the analyses can be helpful to understand the physical mechanism for the unfolding pathway of RNA structures.

## Supporting information

**S1 Text. The force field of the present model, the RNA pseudoknots for 3D structure prediction used in this work, and the 3D structures and stability of RNA pseudoknots predicted by the present model with/without the coaxial stacking potential.**
(PDF)

**S1 Fig. The predicted secondary structure of the six pseudoknots for stability prediction used in this work.** (a) MMTV pseudoknot; (b) T2 pseudoknot; (c) PEMV-1 pseudoknot; (d) BWYV pseudoknot; (e) PLRV pseudoknot; (f) ScYLV pseudoknot.
(TIF)

**S2 Fig. The comparisons between predictions from the present model (solid lines) and other models (dotted lines) for several pseudoknots.** (a,b) BWYV (a) and PEMV-1 (b) pseudoknots at 500mM [$K^+$], respectively. Solid lines: $df/dT$, the first derivative of $f$ with respect to temperature from the present model. Dotted lines: the heat capacity $C_p$ from Ref. 78. Symbols: the heat capacity $C_p$ from experiments [72,73]. (c,d) MMTV pseudoknot at 1000mM [$K^+$] (c) and 50mM [$K^+$] (d), respectively. Solid lines: $df/dT$, the first derivative of $f$ with respect to temperature from the present model. Dotted lines: the heat capacity $C_p$ from Ref. 46. Symbols: the heat capacity $C_p$ from experiments [75].
(TIF)

**S3 Fig. The comparisons between stability of two typical RNA pseudoknots predicted by the present model with and without the coaxial stacking potential.** (a) MMTV pseudoknot at 1000mM [$K^+$]; (b) BWYV pseudoknot at 500mM [$K^+$]. Lines: $df/dT$, the first derivative of $f$ with the temperature; red: predictions from the model with the coaxial stacking potential; green: predictions from the model without the coaxial stacking potential. Cartoon: the predicted 3D structures of the two pseudoknots at different temperatures. Red/black arrow: predictions from the model with coaxial stacking potential; green arrow: predictions from the model without the coaxial stacking potential.
(TIF)

**S4 Fig. The fractions of F, S1, S2 and U states as functions of temperature in thermal unfolding of pseudoknots at 500mM [$Na^+$].** (a) PEMV-1, (b) BWYV, (c) PLRV, and (d) ScYLV pseudoknots. F stands for fully folded RNA; S1, hairpin intermediate with Stem1; S2, hairpin intermediate with Stem2; U, fully unfolded RNA.
(TIF)

**S5 Fig. The folding free energies of the constructs associated with the stems of six wild-type RNA pseudoknots and the variant of T2 pseudoknots at 1M [$Na^+$] as a function of temperature.** (a) MMTV pseudoknot; (b) T2 and T2 variant pseudoknots; (c) PEMV-1 pseudoknot; (d) BWYV pseudoknot; (e) PLRV pseudoknot; and (f) ScYLV pseudoknot. Here, the free energies are computed using Mfold (http://unafold.rna.albany.edu/) [89].
(TIF)

**S6 Fig. The fractions of F, S1, S2 and U states as functions of temperature in thermal unfolding of MMTV and T2 pseudoknots in divalent ion solutions.** (a-c) MMTV pseudoknot at 50mM [$K^+$] and different [$Mg^{2+}$]'s: (a) 0.1mM [$Mg^{2+}$], (b) 1mM [$Mg^{2+}$], and (c) 10mM [$Mg^{2+}$]. (d-f) T2 pseudoknot at 100mM [$K^+$] and different [$Mg^{2+}$]'s: (d) 0.1mM [$Mg^{2+}$], (e) 1mM [$Mg^{2+}$], and (f) 10mM [$Mg^{2+}$].
(TIF)

## Acknowledgments

Parts of the numerical calculation in this work are performed on the super computing system in the Super Computing Center of Wuhan University.

## Author Contributions

**Conceptualization:** Ya-Zhou Shi, Lei Jin, Zhi-Jie Tan.

**Data curation:** Ya-Zhou Shi, Chen-Jie Feng, Ya-Lan Tan, Zhi-Jie Tan.

**Formal analysis:** Ya-Zhou Shi, Ya-Lan Tan, Zhi-Jie Tan.

**Funding acquisition:** Ya-Zhou Shi, Zhi-Jie Tan.

**Investigation:** Ya-Zhou Shi, Chen-Jie Feng.

**Methodology:** Ya-Zhou Shi, Lei Jin, Chen-Jie Feng, Zhi-Jie Tan.

**Project administration:** Zhi-Jie Tan.

**Resources:** Zhi-Jie Tan.

**Supervision:** Zhi-Jie Tan.

**Validation:** Ya-Zhou Shi, Ya-Lan Tan, Zhi-Jie Tan.

**Writing ± original draft:** Ya-Zhou Shi.

**Writing ± review & editing:** Ya-Zhou Shi, Lei Jin, Zhi-Jie Tan.